\documentclass[aps,prb,reprint]{revtex4-1}
\usepackage{amsmath,amssymb,eqnarray}
\usepackage{graphicx}
\usepackage{dcolumn}% Align table columns on decimal point
\usepackage{bm}% bold math
\usepackage{epsfig}
\usepackage{subfigure}
\begin{document}
\title{Cooper Pairing in A Doped 2D Antiferromagnet with Spin-Orbit Coupling}
\author{Jingxiang Zhao$^1$, Qiang Gu$^{1,2}$}
\email[Corresponding author: ]{qgu@ustb.edu.cn}
\affiliation{$^1$Department of Physics, University of Science and Technology Beijing, Beijing 100083, P. R. China\\
$^2$Beijing Key Laboratory for Magneto-Photoelectrical Composite and Interface Science, School of Mathematics and Physics, University of Science and Technology Beijing, Beijing 100083, P. R. China}
\date{\today}

\begin{abstract}
We study the two-dimensional Hubbard model with the Rashba type spin-orbit coupling within and beyond the mean-field theory. The antiferromagnetic ground state for the model at half-filling and the Cooper pairing induced by antiferromagnetic spin fluctuations near half-filling are examined based on the random-phase approximation. We show that the antiferromagnetic order is suppressed and the magnetic susceptibility turns out to be anisotropic in the presence of the spin-orbit coupling. Energy spectrums of transverse spin fluctuations are obtained and the effective interactions between holes mediated by antiferromagnetic spin fluctuations are deduced in the case of low hole doping. It seems that the spin-orbit coupling tends to form s+p-wave Cooper pairs, while the s+d-wave pairing is dominant when the spin-orbit coupling is absent.
\end{abstract}
\maketitle

\section*{I. Introduction}
Spin fluctuations may result in effective attractive interactions between fermions, and this mechanism plays an important role in understanding unconventional superconductivity\cite{Scalapino2012}. And it attracts much attention till today\cite{Essenberger2014,Anderson2016}. For ferromagnetic or nearly ferromagnetic Fermi systems, spin fluctuations are favorable to spin-triplet Cooper pairs. For example, it was suggested that the p-wave triplet Cooper pairing in superfluid $^3$He, a nearly ferromagnetic Fermi liquid, should be induced by spin fluctuations\cite{Osheroff1997}. Furthermore, Fay and Appel pointed out that the longitudinal ferromagnetic spin fluctuations could cause p-wave effective attraction within the ferromagnetic state\cite{Fay1980}. This theory provides an candidate explanation on the superconductivity in the ferromagnetic superconductors, such as UGe$_2$\cite{Saxena2000} and UCoGe\cite{Hattori2012}, whose superconductivity (SC) state coexists with the itinerant-electron ferromagnetic order.

Similarly, antiferromagnetic (AFM) spin fluctuations can also give rise to Cooper pairing. Schrieffer, Wen and Zhang have proposed an AFM spin fluctuation mechanism, the spin-bag model\cite{Schrieffer1989}, to explain the high-T$_c$ superconductivity\cite{Bednorz1986}. This model is based on the half-filled Hubbard model on the square-lattice, which favors the AFM ground state in the large-U limit and thus corresponds to the AFM order of the parent materials of cuprate superconductors. Schrieffer {\it et al.} suggested that the AFM spin fluctuation should induce d-wave Cooper pairs between holes in the weak hole-doping case.

The AFM order is also present in many other superconductors or their parent materials, including heavy-fermion and iron-based superconductors. For instance, the SC phase of the heavy fermion material CePt$_3$Si has been proved to coexist with antiferromagnetism\cite{Bauer2004}, and so has the low-pressure SC phase of CeCu$_2$Si$_2$\cite{Steglich2003}. Various AFM orders appear in iron-based superconductors, such as the collinear AFM state in LaO$_{1-x}$F$_x$FeAs\cite{Dai2008}, the bi-collinear state in Fe$_{1+y}$Se$_x$Te$_{1-x}$\cite{Dai2009}, and the blocked checkerboard AFM order in K$_{0.8}$Fe$_{1.6}$Se$_2$\cite{Xiang2011}. Recently, some groups report the microscopic coexistence of antiferromagnetism and superconductivity in the iron-based materials\cite{Ma2012,Li2012}. It is naturally supposed that AFM spin fluctuations might play an important role in iron-based superconductors. For example, some reports suggest that the anti-ferromagnetic spin fluctuations of LaFeAsO$_{1-x}$F$_x$\cite{Mazin2008} and Fe$_{1-x}$Co$_x$Se\cite{Urata2016} should mediate the s$\pm$ wave superconducting state.

One more important issue is that the spin-orbit coupling (SOC) may be present in some of forgoing superconductors. Particularly, the SOC is inevitably resulted from the lack of structure inversion symmetry and therefore it must be considered in the non-centrosymmetric (NCS) superconductor\cite{Sigrist2004}. It is known that the SOC can cause the admixture of spin-up and spin-down\cite{Zhang2004,Fu2015}, which essentially influences the spin degree of freedom. Resulting from the spin-mixing, SOC could lead to the mixture of spin-singlet and spin-triplet pairing symmetry\cite{Gorkov2001}. Actually, SOC has stimulated much research interest in recent years since it plays an important role in other condensed matter systems, e.g. Quantum spin Hall insulator\cite{Wu2012} and atomic Fermi superfluid\cite{Liao2012}.

Due to the effect of SOC on spin states, it is infered that SOC could affect the spin fluctuations and influence the orbital symmetry of Cooper pairs mediated by spin fluctuations. Various models were studied to discuss the role of SOC in the orbital symmetry. Some papers employed a two-band model and stated that SOC could split the degeneracy of p-wave states\cite{Sigrist2000}. This model combined the itinerant electrons with local moments and could help understand some unconventional superconductivity, e.g., Sr$_2$RuO$_4$. But it was not suitable for some superconductors whose SC and AFM order originated from the same band electrons, for example, some NCS superconductors which also exhibited the AFM order. Therefore, a single-band model was necessary. Some papers studied a single-band Hubbard model with SOC\cite{Yokoyama2007,Shigeta2013} and reported that SOC might be in favor of the d+f-wave pairing states or p+d-wave states. However, these papers neglected the AFM fluctuation. In some intensive studies on NCS superconductivity, the AFM fluctuation was introduced in different ways. Some people selected parameters to fit spin susceptibility into experimental results manifesting anti-ferromagnetism\cite{Tada2008}, while some people employed a staggered field to describe the AFM order\cite{Yanase2008}. All of them reported that the mixture of spin-singlet and triplet Cooper pairs was resulted from SOC and the orbital symmetry was obtained, e.g., s+p-wave or p+d+f-wave states. But the AFM order could not consistently obtained in these papers. Thus a better single-band model is necessary for consistently studying the effect of SOC on the AFM order and fluctuations.

In this paper, we investigate the half-filled Hubbard model with the Rashba SOC in a two-dimensional (2D) square lattice. The central issue of this paper is to examine the influence of SOC on the Cooper pairing intermediated by the AFM spin fluctuations. Our model is a single-band model, which suggests both AFM order and superconductivity are originated from one-band electrons. It is might help for understanding the magnetic properties and pairing symmetry of some quasi two-dimensional layered superconductors with SOC, for instance, iron-based superconductors\cite{Kamihara2008,Rotter2008} and NCS\cite{Pereiro2011}.

The paper is organized as follows. The model is described in Section II. Ground state properties for the model at half-filled are studied based on the mean-field approximation and the RPA. The sublattice magnetization, the spectrum of transverse spin excitation and the ratio of transverse versus longitudinal spin susceptibility at $(\pi,\pi)$ are calculated. Section IV discusses effective interactions between holes induced by the AFM spin fluctuations in the case of weak hole doping, with the emphasis on pairing effects in the $s$, $p$, and $d$ channels. The conclusions are given in the last section.

\section*{II. The mean-field model}
We start from a single-band half-filled Hubbard model with SOC in a two-dimensional square lattice.
\begin{align}
H=\sum_{k,\sigma}\varepsilon_kc_{k\sigma}^{\dagger}c_{k\sigma}+\alpha\sum_k g(\vec{k})\cdot s(\vec{k})+U\sum_{i}n_{i\uparrow}n_{i\downarrow}~,\nonumber
\end{align}
where $\varepsilon_k=-2t(\cos{k_xa}+\cos{k_ya})$ is the kinetic energy arising from electron hoping between the nearest neighbours with $a$ being the lattice constant. The second term of Hamiltonian describes the spin-orbit coupling with $\alpha$ being the coupling strength.
Here the type of SOC takes the form as\cite{Yanase2008,Yokoyama2007}:
$s(\vec{k})$=$\sum_{\sigma,\sigma^\prime}\sigma_{\sigma,\sigma^\prime} c^{\dagger}_{k,\sigma}c_{k,\sigma^\prime}$ and
$g(\vec{k})$=$(-v_y(\vec{k}), v_x(\vec{k}),0)$, where $v_{x,y}(\vec{k})$=$\partial \varepsilon_k/\partial k_{x,y}$. With the
definition, the $g$ vector, which is $(-2t\sin k_ya, 2t\sin k_xa,0)$,
protects the symmetry and periodicity of the Brillouin zone. In the following, we assume $a=1$ for simplicity.

By introducing $v(\vec{k})$=$-2t\sqrt{\sin^2 k_y+\sin^2 k_x}$, the term of SOC,
$-2t\alpha\sum_k(\sin k_y\pm i\sin k_x)$ can be denoted as
$\alpha\sum_k v(\hat{k})\exp(\pm i \phi_k)$, where
$\phi_k$=$\arctan \left(\sin k_x/\sin k_y\right)$. In this case, the Hamiltonian has the form,
\begin{align}
H&=\sum_{k,\sigma}\varepsilon_kc^{\dagger}_{k\sigma}c_{k\sigma}
+\frac{U}{2N}\sum_{k,k^{\prime},q}\sum_{\sigma\sigma^\prime}c^{\dagger}_{k-q\sigma}c^{\dagger}_{k^{\prime}+q\sigma^\prime}
c_{k^{\prime}\sigma^\prime}c_{k\sigma}\nonumber\\
&+\alpha\sum_k v(\hat{k})e^{i \phi_k}c_{k\uparrow}^{\dagger}c_{k\downarrow}+
\alpha\sum_kv(\hat{k})e^{-i \phi_k}c_{k\downarrow}^{\dagger}c_{k\uparrow}~.
\end{align}

\begin{figure}
    \centering
    \includegraphics[width=0.4\textwidth]{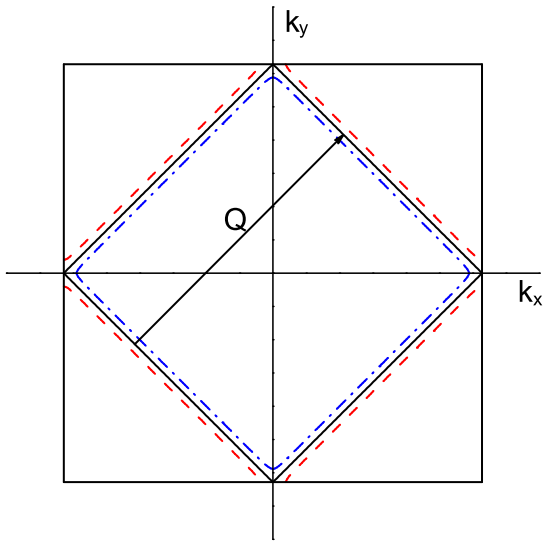}
    \caption{(Color online) The schematic of the first Brillouin zone; the solid line represents the Fermi surface at half-filling without SOC. The dashed lines represent the split Fermi surfaces resulted from SOC.}\label{FS}
\end{figure}

\begin{figure}
    \centering
    \includegraphics[width=0.4\textwidth]{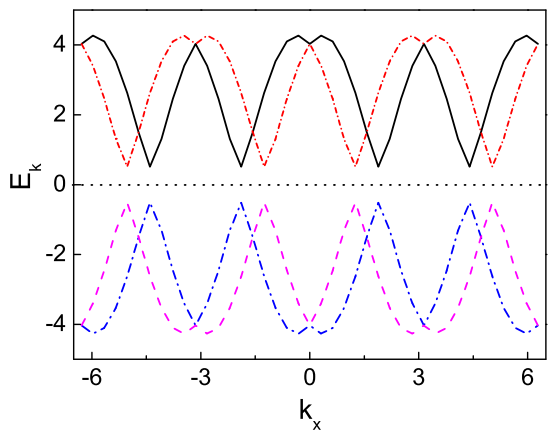}
    \caption{(Color online) The schematic diagram of the eigenvalues. It is shown that the period of energy bands are protected by Rashba SOC. The dotted line locating in the middle of figure represents the chemical potential.}\label{E}
\end{figure}

In the case of AFM, by using the mean-field approach (MFA), the interaction term can be written as
$-US\sum_{k,\sigma,\sigma^\prime}c^{\dagger}_{k+Q,\sigma}\sigma^z_{\sigma\sigma^\prime}c_{k,\sigma^\prime}$\cite{Schrieffer1989,Anderson2016},
where $Q=(\pi,\pi)$, the nesting vector of Fermi surface as shown in Fig.~\ref{FS}, and $S=\left<G\left|S_Q^z\right|G\right> / N$, where $S_Q^z=\sum_{k}
c^{\dagger}_{k+Q,\sigma}\sigma^z_{\sigma\sigma^\prime}c_{k,\sigma^\prime}$ and
$\left|G\right>$ is the ground state of the model. Therefore, the AFM order can be studied consistently. When the spin-orbit
coupling is ignored, $\left|G\right>$ is the same as the ground state of
the antiferromagnetism defined as in Ref.[\onlinecite{Schrieffer1989}].
Through introducing new fermion-operators $f_{k,\eta}$ ($\eta=1,2,3,4$), the Hamiltonian can be
diagonalized via the Bogliubov transformation. In the process of diagonalization, some equations between $k$ and $k+Q$
are: $i)$ the nesting Fermi surface results in $\varepsilon_{k+Q}=-\varepsilon_k$; $ii)$ the principal value of $\phi_k$ is confined in $(-\pi, \pi]$. Accordingly $\phi_{k+Q}$=$\phi_k+\pi$ is defined to keep $v(k+Q)e^{-i \phi_{k+Q}}$=$-v(k) e^{-i k\phi_k}$ and $\phi_{-k}=\phi_k+\pi$ to preserve $g(-k)=-g(k)$.
The relationship between electron operators and the quasi-particles operators is expressed as,
\begin{align}
\label{Transformation}
&c_{k\uparrow}=e^{\frac{i\phi_k}{2}}\left[u_{k+}f^c_{k,+}+u_{k-}f^c_{k,-}+\nu_{k+}f^v_{k,+}+\nu_{k-}f^v_{k,-}\right]~,\nonumber\\
&c_{k\downarrow}=e^{\frac{-i\phi_k}{2}}\left[-u_{k+}f^c_{k,+}+u_{k-}f^c_{k,-}-\nu_{k+}f^v_{k,+}+\nu_{k-}f^v_{k,-}\right]~,\nonumber\\
&c_{k+Q\uparrow}=e^{\frac{i\phi_k}{2}}\left[\nu_{k+}f^c_{k,+}+\nu_{k-}f^c_{k,-}-u_{k+}f^v_{k,+}-u_{k-}f^v_{k,-}\right]~,\nonumber\\
&c_{k+Q\downarrow}=e^{\frac{-i\phi_k}{2}}\left[\nu_{k+}f^c_{k,+}-\nu_{k-}f^c_{k,-}-u_{k+}f^v_{k,+}+u_{k-}f^v_{k,-}\right]~,
\end{align}
where, $u_{\pm}=\frac{1}{2}\sqrt{1+\frac{\xi_{k,\pm}}{E_{k,\pm}}}$, $\nu_{\pm}=\frac{1}{2}\sqrt{1-\frac{\xi_{k,\pm}}{E_{k,\pm}}}$, $E_{k,\pm}=\sqrt{\xi_{k,\pm}^2+\Delta^2}$ is the eigenvalue with $\xi_{k,\pm}=\varepsilon_k\pm\alpha v_k$ and $\Delta=-US/2$.
The diagonalized Hamiltonian can be formed as below,
\begin{align}
H=&\sum_k\,^\prime E_{k,+}\left(f_{k,+}^{c\dagger}f^c_{k,+}-f_{k,+}^{v\dagger}f^v_{k,+}\right)\nonumber\\
&+\sum_k\,^\prime E_{k,-}\left(f_{k,-}^{c\dagger}f^c_{k,-}-f_{k,-}^{v\dagger}f^v_{k,-}\right)~.
\end{align}
where $\sum^\prime$ represents the summation extending over the magnetic zone without SOC displayed in Fig.~\ref{FS}. The Fermi surface and the energy spectrum of electrons are split by SOC as shown in Fig.~\ref{FS} and Fig.~\ref{E}, respectively. The nesting Fermi surface which stands for the AFM order is broken by SOC, which suggests that SOC should suppress AFM order. Moreover, the system still retains the periodicity as shown in Fig.~\ref{E} so the first BZ can still represent the symmetry of the model. So the numerical analysis will be reduced in the first BZ rather than magnetic BZ in the presence of SOC.


In the half-filled case, the conductive bands,
$f^c_{k,+}$ and $f^c_{k,-}$, are empty for particles and the other
bands, $f^v_{k,+}$ and $f^v_{k,-}$ are valence bands which are
full filled by particles, so the ground state can be defined as:
\begin{align}
\label{G_state}
f^c_{k,+}\left|G\right>=f^c_{k,-}\left|G\right>=0~;f_{k,+}^{v\dagger}\left|G\right>=f_{k,-}^{v\dagger}\left|G\right>=0~.
\end{align}

\section*{III. Ground state properties on the random-phase approximation}
\begin{figure}
  \centering
    \includegraphics[width=0.4\textwidth]{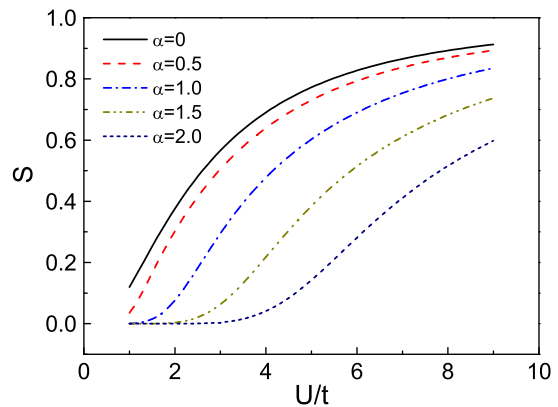}
    \caption{(Color online) The sublattice magnetization S of the system with SOC obtained by the mean-field approach. $\alpha$ is the strength of the reduced spin-orbit coupling.}\label{S_on_U_with_soc_mf}
\end{figure}

\begin{figure}
  \centering
  \includegraphics[width=0.4\textwidth]{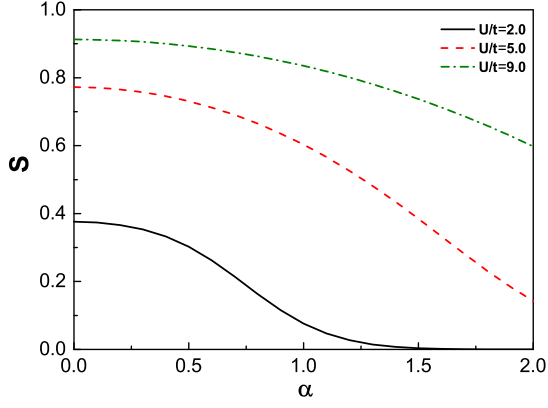}
  \caption{(Color online) The sublattice magnetization S v.s. the strength of SOC $\alpha$ for different Hubbard interaction U, which is obtained by mean-field approach.}\label{S_SOC}
\end{figure}

To quantitatively study the effect exerted by SOC on the AFM order, we employ the foregoing definitions of ground state to obtain the self-consistent equation of the sublattice magnetization $S$:
\begin{align}
\label{MF}
S&=\frac{1}{N}\left<G\left|S_Q^z\right|G\right>\nonumber\\
&=-\frac{1}{N}\sum_k\,^\prime\left(\frac{\Delta}{E_{k,+}}+\frac{\Delta}{E_{k,-}}\right)=-\frac{2\Delta}{U}~.
\end{align}
We show the relationship between $S$ and the Hubbard interaction $U/t$ in Fig.~\ref{S_on_U_with_soc_mf}. It must be pointed out that the MFA is more applicable when the Hubbard interaction is strong, so U/t$>$1 is shown. It does not imply that a critical value of Hubbard interaction is defined. As shown, $S$ increases as the Hubbard interaction $U$ is enhanced. To study the role of SOC theoretically, the strength of SOC is selected from 0 to 2. The results with $\alpha=0$ correspond to the absence of SOC\cite{Schrieffer1989}. If $U$ is fixed, we can find that $S$ decreases with $\alpha$ increased. Fig.~\ref{S_SOC} exhibits that the weaker the interaction $U$ is, the smaller SOC suppressing the magnetization to zero is. These results suggest that the Hubbard interaction be beneficial to AFM order, while AFM order should be suppressed by SOC.
The suppression of $S$ by SOC might be due to the width-broadening of the Hubbard bands by SOC. It is similar to the decreasing of Hubbard interaction. The system might be in favor of paramagnetic metal when U is small\cite{Chen2012}. As shown in Fig.~\ref{S_on_U_with_soc_mf}, the weaker $U$ is, the smaller $S$ is.

It is known that the MF approach can give a qualitative description. To quantitatively study the effect of sublattice on the Hubbard interaction U, We have to discuss the effect of fluctuations on the ground state. Based on the ground state of the model, the definitions of charge and spin correlation functions of electrons are
Ref.~[\onlinecite{Schrieffer1989}],
\begin{align}\label{CF}
\bar{\chi}^{00}(q,q^\prime,t)&=\frac{i}{2N}\left<G\right|\mathcal{T}\rho_q(t)\rho_{-q^\prime}(0)\left|G\right>\nonumber \\
\bar{\chi}^{ij}(q,q^\prime,t)&=\frac{i}{2N}\left<G\right|\mathcal{T}S^i_q(t)S^j_{-q^\prime}(0)\left|G\right>~,
\end{align}
where $\rho_q=\sum_{k,\alpha}c_{k+q,\alpha}^{\dagger}c_{k,\alpha}$ is the charge density operator, and $S^{i}_{q}=\sum_{k,\alpha,\beta}c^{\dagger}_{k+q,\alpha}\sigma^i_{\alpha\beta}c_{k,\beta}$ is the spin density operator. $\left|G\right>$ is the ground state defined by Eq.~(\ref{G_state}). In the case of half-filling, the non-vanishing terms of Eq.~(\ref{CF}) are
\begin{align}
\left<f^{v\dagger}_{p\lambda}(t_1)f^{c}_{k\lambda^\prime}(t_1)f^{c\dagger}_{k\lambda^\prime}(t)f^{v}_{p\lambda}(t)\right>~,
\end{align}
where, $t<t_1$. The reason lies in that a particle only annihilates firstly, then creates in the fulled valence bands and the process is just the opposite one in the empty conduction bands.

Based on the transformation, Eq.~(\ref{Transformation}), the correlation functions with SOC can be obtained as below,

\begin{widetext}
\begin{align}
\bar{\chi}^{00}_0(q,\omega)=&\sum_{k,\sigma}G^0_{\sigma\sigma}(k+q,\omega)G^0_{\sigma\sigma}(k+q,\omega)
+\sum_kG^0_{\uparrow\downarrow}(k+q,\omega)G^0_{\downarrow\uparrow}(k+q,\omega)+\sum_kG^0_{\downarrow\uparrow}(k+q,\omega)G^0_{\uparrow\downarrow}(k+q,\omega)\nonumber\\
=&-\frac{1}{2N}\sum_k\,^\prime\left(1+e^{-i(\phi_{k+q}-\phi_k)}+e^{i(\phi_{k+q}-\phi_k)}\right)\left[\bar{\chi}^{00}_{0f1}(q,\omega)+\bar{\chi}^{00}_{0f2}(q,\omega)
+\bar{\chi}^{00}_{0f3}(q,\omega)+\bar{\chi}^{00}_{0f4}(q,\omega)\right]~.
\end{align}
\begin{align}\label{longcr}
&\bar{\chi}^{00}_{0f1}(q,\omega)=\frac{1}{4}\left[\left(1-\frac{\xi_{k,+}\xi_{k+q,+}+\Delta^2}{E_{k,+}E_{k+q,+}}\right)
\left(\frac{1}{\omega-E_{k,+}-E_{k+q,+}+i\delta}+\frac{1}{-\omega-E_{k,+}-E_{k+q,+}+i\delta}\right)\right]\nonumber\\
&\bar{\chi}^{00}_{0f2}(q,\omega)=\frac{1}{4}\left[\left(1-\frac{\xi_{k,+}\xi_{k+q,-}+\Delta^2}{E_{k,+}E_{k+q,-}}\right)
\left(\frac{1}{\omega-E_{k,+}-E_{k+q,-}+i\delta}+\frac{1}{-\omega-E_{k,+}-E_{k+q,-}+i\delta}\right)\right]\nonumber\\
&\bar{\chi}^{00}_{0f3}(q,\omega)=\frac{1}{4}\left[\left(1-\frac{\xi_{k,-}\xi_{k+q,+}+\Delta^2}{E_{k,-}E_{k+q,+}}\right)
\left(\frac{1}{\omega-E_{k,-}-E_{k+q,+}+i\delta}+\frac{1}{-\omega-E_{k,-}-E_{k+q,+}+i\delta}\right)\right]\nonumber\\
&\bar{\chi}^{00}_{0f4}(q,\omega)=\frac{1}{4}\left[\left(1-\frac{\xi_{k,-}\xi_{k+q,-}+\Delta^2}{E_{k,-}E_{k+q,-}}\right)
\left(\frac{1}{\omega-E_{k,-}-E_{k+q,-}+i\delta}+\frac{1}{-\omega-E_{k,-}-E_{k+q,-}+i\delta}\right)\right]~.
\end{align}
The singularity of $\bar{\chi}^{00}_{0fi}$ is in the lower half plane when $t>0$. When $t<0$, the singularity is in the upper half plane.
The longitudinal spin correlation function,
\begin{align}
\bar{\chi}^{zz}_0(q,\omega)=&\sum_{\sigma}G^0_{\sigma\sigma}(k+q,\omega)G^0_{\sigma\sigma}(k+q,\omega)
-\sum_kG^0_{\uparrow\downarrow}(k+q,\omega)G^0_{\downarrow\uparrow}(k+q,\omega)-\sum_kG^0_{\downarrow\uparrow}(k+q,\omega)G^0_{\uparrow\downarrow}(k+q,\omega)\nonumber\\
&=-\frac{1}{2N}\sum_k\,^\prime\left(1-e^{-i(\phi_{k+q}-\phi_k)}-e^{i(\phi_{k+q}-\phi_k)}\right)\left[\bar{\chi}^{00}_{0f1}(q,\omega)+\bar{\chi}^{00}_{0f2}(q,\omega)
+\bar{\chi}^{00}_{0f3}(q,\omega)+\bar{\chi}^{00}_{0f4}(q,\omega)\right]~.
\end{align}
We can find that $\bar{\chi}^{zz}_0(q,\omega)$ is different from $\bar{\chi}^{00}_0(q,\omega)$, other than the case without SOC, where $\bar{\chi}^{zz}_0(q,\omega)$=$\bar{\chi}^{00}_0(q,\omega)$. This is because $G^0_{\uparrow\downarrow}$ and $G^0_{\downarrow\uparrow}$ are non-zero as the spins mixed by SOC.
The transverse spin correlation function are presented as follows,
\begin{align}
&\bar{\chi}^{+-}_o(q,\omega)=-\frac{1}{2N}\sum_k\,^\prime\left[\bar{\chi}^{+-}_{of1}(q,\omega)+\bar{\chi}^{+-}_{of2}(q,\omega)
+\bar{\chi}^{+-}_{of3}(q,\omega)+\bar{\chi}^{+-}_{of4}(q,\omega)\right]\\
&\bar{\chi}^{+-}_Q(q,\omega)=-\frac{1}{2N}\sum_k\,^\prime\left[\bar{\chi}^{+-}_{Qf1}(q,\omega)+\bar{\chi}^{+-}_{Qf2}(q,\omega)
+\bar{\chi}^{+-}_{Qf3}(q,\omega)+\bar{\chi}^{+-}_{Qf4}(q,\omega)\right]~,
\end{align}
\begin{align}
\label{transversecr1}
&\bar{\chi}^{+-}_{of1}(q,\omega)=\frac{1}{4}\left[\left(1-\frac{\xi_{k,+}\xi_{k+q,+}-\Delta^2}{E_{k,+}E_{k+q,+}}\right)
\left(\frac{1}{\omega-E_{k,+}-E_{k+q,+}+i\delta}+\frac{1}{-\omega-E_{k,+}-E_{k+q,+}+i\delta}\right)\right]\nonumber\\
&\bar{\chi}^{+-}_{of2}(q,\omega)=\frac{1}{4}\left[\left(1-\frac{\xi_{k,+}\xi_{k+q,-}-\Delta^2}{E_{k,+}E_{k+q,-}}\right)
\left(\frac{1}{\omega-E_{k,+}-E_{k+q,-}+i\delta}+\frac{1}{-\omega-E_{k,+}-E_{k+q,-}+i\delta}\right)\right]\nonumber\\
&\bar{\chi}^{+-}_{of3}(q,\omega)=\frac{1}{4}\left[\left(1-\frac{\xi_{k,-}\xi_{k+q,+}-\Delta^2}{E_{k,-}E_{k+q,+}}\right)
\left(\frac{1}{\omega-E_{k,-}-E_{k+q,+}+i\delta}+\frac{1}{-\omega-E_{k,-}-E_{k+q,+}+i\delta}\right)\right]\nonumber\\
&\bar{\chi}^{+-}_{of4}(q,\omega)=\frac{1}{4}\left[\left(1-\frac{\xi_{k,-}\xi_{k+q,-}-\Delta^2}{E_{k,-}E_{k+q,-}}\right)
\left(\frac{1}{\omega-E_{k,-}-E_{k+q,-}+i\delta}+\frac{1}{-\omega-E_{k,-}-E_{k+q,-}+i\delta}\right)\right]~,
\end{align}
\begin{align}
\label{transversecr2}
&\bar{\chi}^{+-}_{Qf1}(q,\omega)=\frac{\Delta}{E_{k,+}}
\left(\frac{1}{\omega-E_{k,+}-E_{k+q,+}+i\delta}-\frac{1}{-\omega-E_{k,+}-E_{k+q,+}+i\delta}\right)\nonumber\\
&\bar{\chi}^{+-}_{Qf2}(q,\omega)=\frac{\Delta}{E_{k,+}}
\left(\frac{1}{\omega-E_{k,+}-E_{k+q,-}+i\delta}-\frac{1}{-\omega-E_{k,+}-E_{k+q,-}+i\delta}\right)\nonumber\\
&\bar{\chi}^{+-}_{Qf3}(q,\omega)=\frac{\Delta}{E_{k,-}}
\left(\frac{1}{\omega-E_{k,-}-E_{k+q,+}+i\delta}-\frac{1}{-\omega-E_{k,-}-E_{k+q,+}+i\delta}\right)\nonumber\\
&\bar{\chi}^{+-}_{Qf4}(q,\omega)=\frac{\Delta}{E_{k,-}}
\left(\frac{1}{\omega-E_{k,-}-E_{k+q,-}+i\delta}-\frac{1}{-\omega-E_{k,-}-E_{k+q,-}+i\delta}\right)~.
\end{align}
\end{widetext}
In the limit of $\alpha=0$, all the correlation functions are the same as Ref.~(\onlinecite{Schrieffer1989}). The subscript $o$ and $Q$ in Eq.~(\ref{transversecr1}) and (\ref{transversecr2}) are $q^\prime=q$ and $q^\prime=q+Q$ respectively.
\begin{align}
\chi_0^{+-}(q,q^\prime;\omega)=&\chi_o^{+-}(q;\omega)\delta(q^\prime-q)\nonumber\\
&+\chi_Q^{+-}(q,\omega)\delta(q^\prime-q+Q)~,
\end{align}
is a 2 $\times$ 2 matrix.

\begin{figure}
  \centering
  \includegraphics[width=0.4\textwidth]{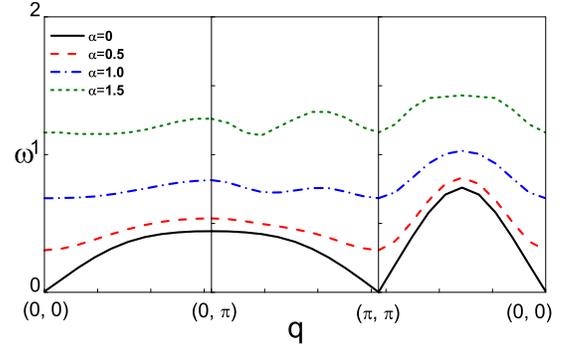}
  \caption{(Color online)The energy spectrum of transverse spin fluctuation with different strength of SOC, which is along the symmetry route (0 ,0) $\rightarrow$ (0, $\pi$) $\rightarrow$ ($\pi$, $\pi$) $\rightarrow$ (0, 0).}\label{sus}
\end{figure}

With RPA, the susceptibility is obtained by the following formulas,
\begin{align}
\label{chargesus}
&\bar{\chi}_{RPA}^{00}(q,\omega)=\sum_{i}\frac{\bar{\chi}^{00}_{0fi}(q,\omega)}{1+U\bar{\chi}^{00}_{0fi}(q,\omega)}~,\\
\label{longsus}
&\bar{\chi}_{RPA}^{zz}(q,\omega)=\sum_{i}\frac{\bar{\chi}^{zz}_0(q,\omega)}{1-U\bar{\chi}^{zz}_0(q,\omega)}~,\\
\label{transesus}
&\bar{\chi}_{RPA}^{+-}(q,\omega)\nonumber\\
&\quad=\sum_{i}\sum_{q^\prime}\chi_{ofi}^{+-}(q,q^\prime,\omega)(1-U\chi_{0fi}^{+-}(q,q^\prime,\omega))^{-1}~.
\end{align}
To investigate the effect of SOC on the spin fluctuations, the energy spectrum of transverse spin fluctuations is deduced by calculating the pole of transverse dynamical spin susceptibility, as shown in Fig.~\ref{sus}. In the absence of SOC, there is a gapless point at q=($\pi$, $\pi$), which is consistent with the Goldstone theorem\cite{Goldstone1962}. However, a gap opens in the presence of SOC, even though the AFM order remains. The similar phenomenon is also reported in Ref.~[\onlinecite{Nagao2014}] which have discussed the magnetic excitation in Sr$_2$IrO$_4$. A gap will open at the (0, 0) point, which is due to the spin-orbit coupling. That the gap opens in the presence of SOC might result from that SOC breaks the continuous symmetry\cite{Kim2015}, so Goldstone theory is not applicable.

\begin{figure}[h!]
  \centering
  \includegraphics[width=0.4\textwidth]{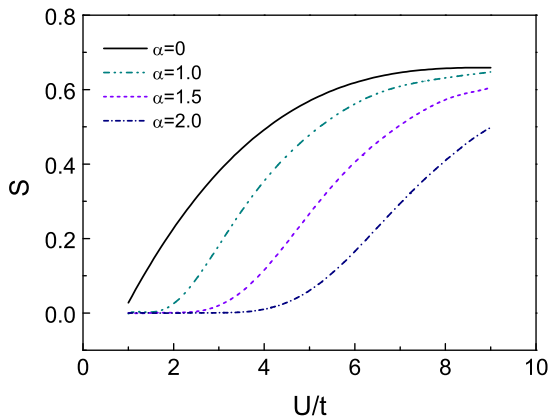}
  \caption{(Color online) The sublattice magnetization S is affected by fluctuations for different SOC, which are obtained by RPA.}\label{m_rpa}
\end{figure}

To study the effect of the fluctuations on the sublattice magnetization, we use the definition of sublattice magnetization according to Ref.~[\onlinecite{Schrieffer1989}],
\begin{align}
S=-\frac{i}{N}\sum_k\,^\prime\int\frac{d\omega}{2\pi} {\bf Tr} \left[\sigma^z G^0(k,k+Q;\omega)\right]
\end{align}
With ignoring fluctuations, we define the single Green function with respect to the ground state, $\left|G\right>$,
\begin{equation}
G^0(k,k+Q;\omega)=\frac{\Delta \sigma^z}{\omega^2-E^2_{k+}+i\delta}+\frac{\Delta \sigma^z}{\omega^2-E^2_{k-}+i\delta}~.
\end{equation}
In this case,
\begin{align}
S=&-\frac{1}{N}\sum_k\,^\prime\left(\frac{\Delta}{E_{k+}}+\frac{\Delta}{E_{k-}}\right)\nonumber\\
=&-\frac{2\Delta}{U}~,\label{sg}
\end{align}
which is the same as the result of the MFA,
Eq.~(\ref{MF}). When the fluctuations are considered in our
model, the full Green's function can be obtained by Dyson's equation
with the self-energy which is established by
Eq.~(\ref{chargesus}-\ref{transesus}).
\begin{align}
\frac{1}{G_{\alpha\beta}(k,k^\prime;\omega)}=\frac{1}{G^0_{\alpha\beta}(k,k^\prime;\omega)}-\Sigma_{\alpha\beta}(k,k^\prime;\omega)~,
\end{align}
\begin{figure}[htbp]
  \centering
  \includegraphics[width=0.4\textwidth]{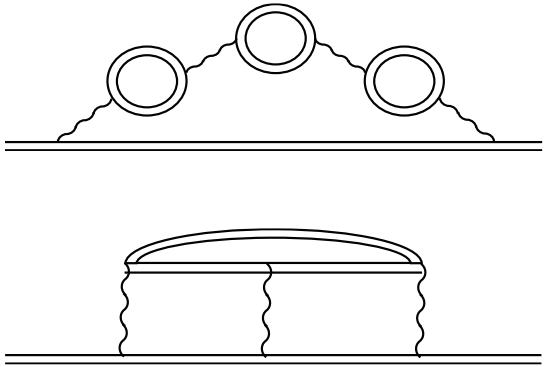}
  \caption{The Feynman diagram of Green function modified by longitudinal spin-fluctuations and transverse spin-fluctuations. The double lines are the single-particle Green function of the quasi-particles}\label{gfzz}
\end{figure}
where $k^\prime=k+Q$. Replacing $G^0$ by the full Green function $G$ in Eq.(\ref{sg}), the numerical results of the sublattice magnetization is shown in Fig.~\ref{m_rpa}. The magnetization is suppressed by the SOC, which agrees with the results obtained by MFA. Comparing with the MFA, the value of $S$ is smaller with the same strength of SOC.

\begin{figure}[htbp]
  \centering
  \includegraphics[width=0.4\textwidth]{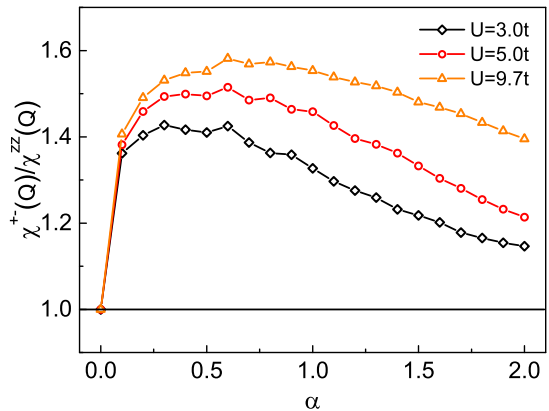}
  \caption{(Color online) The ratio of the transverse and longitudinal susceptibility for different Hubbard interactions. The triangle, circle and rhombus are together on "1" when the strength of SOC is zero.}\label{ratio}
\end{figure}

The ratio of the transverse susceptibility to the longitudinal one with $\bf{q}=(\pi,\pi)$ is calculated. The dependence of ratio on the SOC is exhibited by Fig.~\ref{ratio}. The Hubbard interaction $U=9.7t$ is selected to correspond with Ref.~[\onlinecite{Li2011}], which suggests that the anisotropy of antiferromagnetic fluctuations of Ba$_{0.68}$K$_{0.32}$Fe$_2$As$_2$ may be due to the spin-orbit coupling. The parent material of Ba$_{0.68}$K$_{0.32}$Fe$_2$As$_2$ emerges AFM order and can be regraded as quasi two-dimensional square lattice\cite{Rotter2008}, which can be illuminated via our model. The role of SOC in the anisotropy of susceptibility for Sr2RuO4 was also reported by Eremin {\it et al.}\cite{Eremin2002}, which is agreed with our study. Our calculation shows that $1)$ the ratio is 1 when $\alpha$=0, that is the spin fluctuations are isotropic in the absence of the spin-orbit coupling; $2)$ the ratio is larger than 1 with the increasing of the coupling, which indicates that the spin-orbit coupling can result in an anisotropy of the spin-fluctuations which agrees with Ref.~[\onlinecite{Li2011}]; $3)$ ratio decreases with the Hubbard interaction decreased when we fix the strength of coupling.

\section*{IV. Cooper pairing near half-filling}
In the absence of SOC, the anti-ferromagnetic fluctuations of the weak holes doped Hubbard model favors d-wave paring, which has been used to give an interpretation on the cuprates superconductors\cite{Schrieffer1989}. In this section, the effective interaction intermediated by anti-ferromagnetic spin fluctuations with SOC is studied and the symmetry of Cooper pairs is discussed. We now assume that the system is weak holes doped. At half-filling, the valence band is full and the conduction band is empty. The electrons of the top of the valence band are removed in the case of weak hole doping, and the system can emerge a metallic behavior and superconductivity resulted from the holes. First, the BCS type Hamiltonian is,
\begin{align}
H_{int}&=\frac{1}{4N}\sum_{k,k^\prime}\sum_{
\mbox{
\tiny
$\begin{array}{c}
\alpha\alpha^\prime\\
\beta\beta^\prime
\end{array}
$
}}V_{c}(k^\prime,k)\delta_{\alpha^\prime\alpha}\delta_{\beta^\prime\beta}c^{\dag}_{k^\prime\alpha^\prime}
c^{\dag}_{-k^\prime\beta^\prime}c_{-k\beta}c_{k\alpha}\nonumber\\
&-\frac{1}{4N}\sum_{k,k^\prime}\sum_{
\mbox{
\tiny
$\begin{array}{c}
\alpha\alpha^\prime\\
\beta\beta^\prime
\end{array}
$
}}V_{z}(k^\prime,k)\sigma^z_{\alpha^\prime\alpha}\sigma^z_{\beta^\prime\beta}c^{\dag}_{k^\prime\alpha^\prime}
c^{\dag}_{-k^\prime\beta^\prime}c_{-k\beta}c_{k\alpha}\nonumber\\
&-\frac{1}{4N}\sum_{k,k^\prime}\sum_{
\mbox{
\tiny
$\begin{array}{c}
\alpha\alpha^\prime\\
\beta\beta^\prime
\end{array}
$
}}V_{+-}(k^\prime,k)\sigma^+_{\alpha^\prime\alpha}\sigma^-_{\beta^\prime\beta}c^{\dag}_{k^\prime\alpha^\prime}
c^{\dag}_{-k^\prime\beta^\prime}c_{-k\beta}c_{k\alpha}~.
\end{align}
where $V^c(k^\prime,k)$ is induced by charge-fluctuations and $V^z(k^\prime,k)$ and $V^{+-}(k^\prime,k)$ are caused by longitudinal and transverse fluctuations, respectively.
\begin{align}
&V_{c}(k^\prime,k)=2U-\bar{\chi}_{RPA}^{00}(k^\prime,k)~,\nonumber\\
&V_{z}(k^\prime,k)=\bar{\chi}_{RPA}^{zz}(k^\prime,k)~,\nonumber\\
&V_{+-}(k^\prime,k)=\bar{\chi}_{RPA}^{+-}(k^\prime,k)~.
\end{align}
\begin{figure*}
\centering
\subfigure{
    \begin{minipage}{5cm}
    \centering
    \includegraphics[width=5cm]{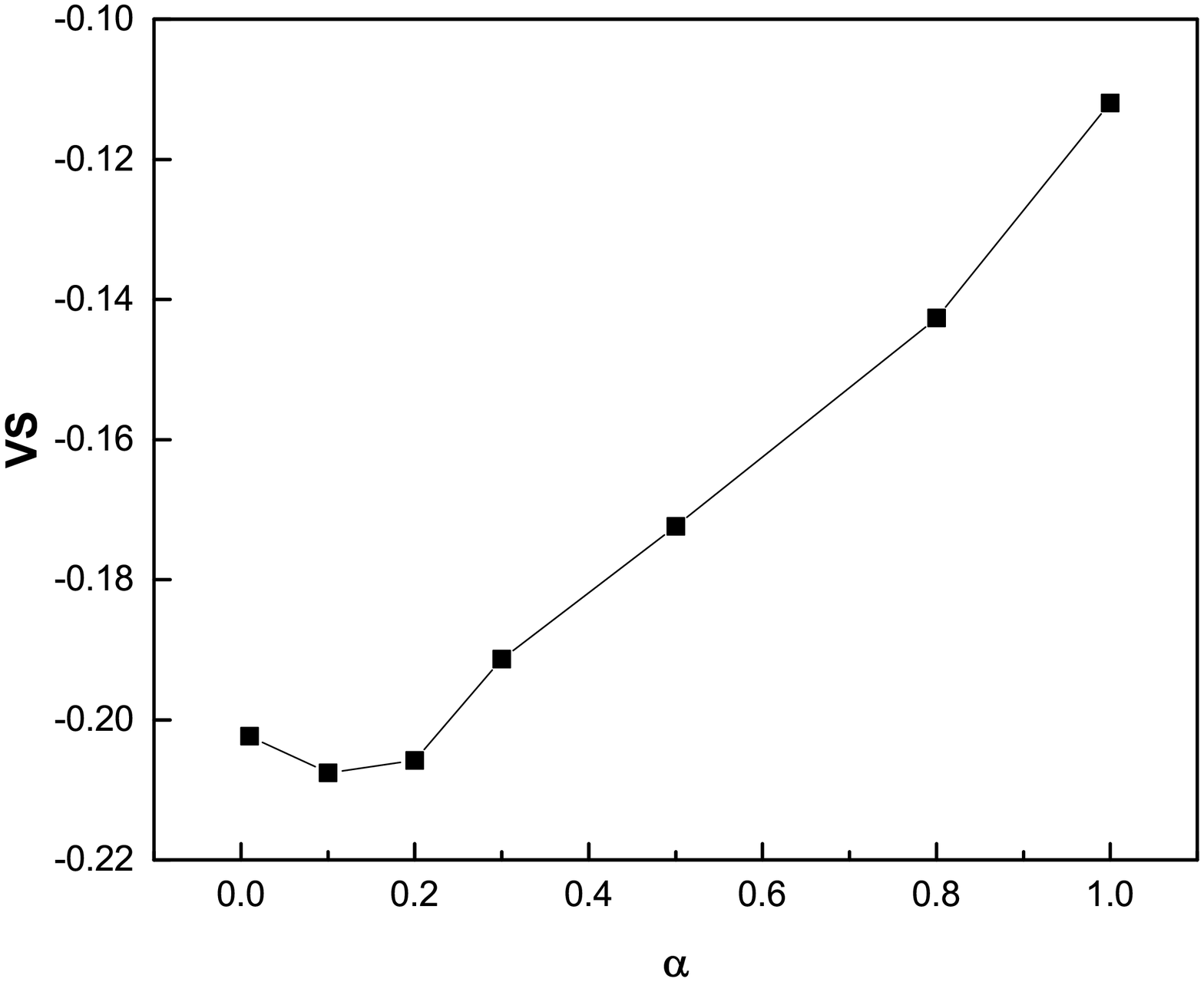}
    \end{minipage}}
\subfigure{
    \begin{minipage}{5cm}
    \centering
    \includegraphics[width=5cm]{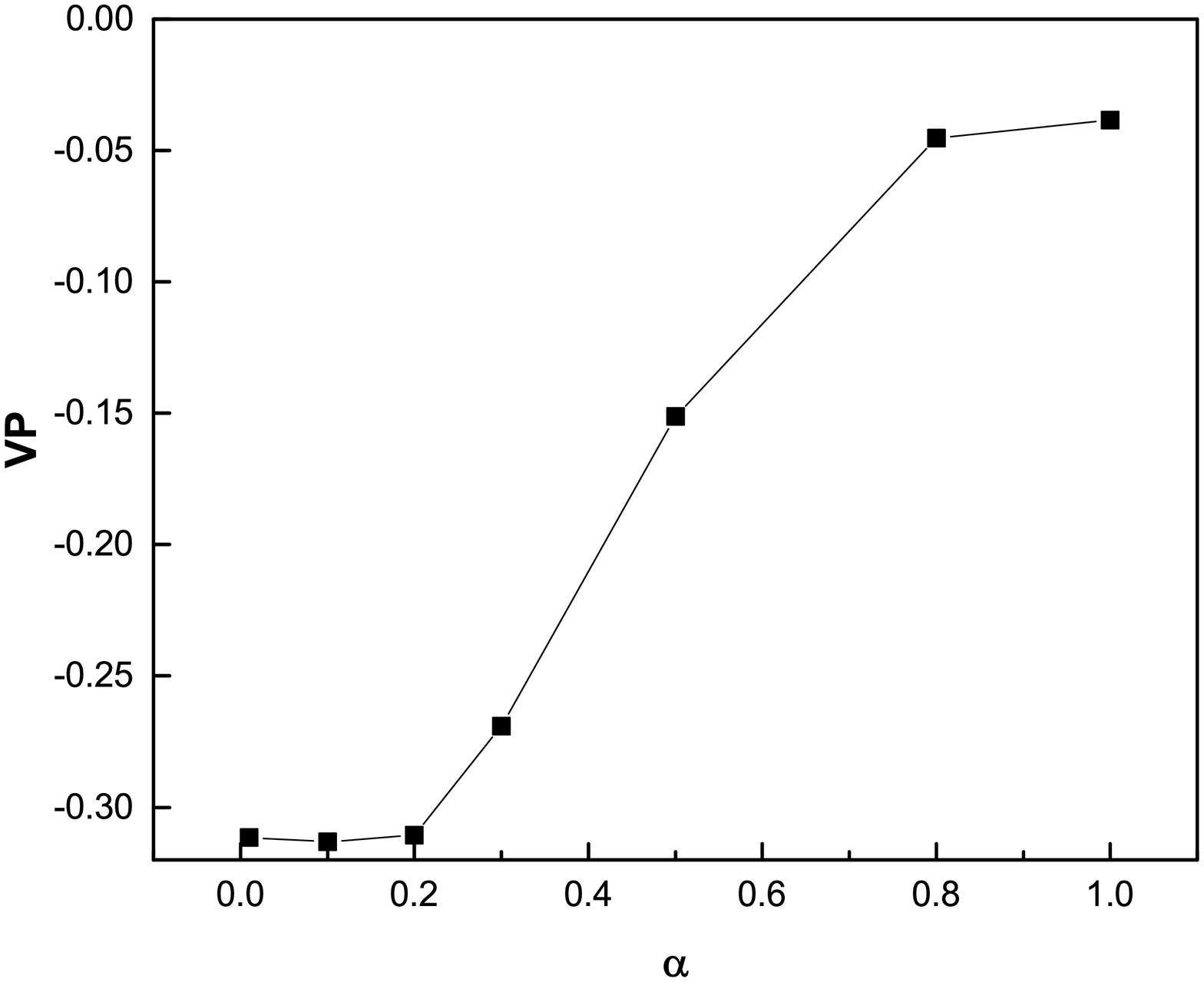}
    \end{minipage}}
\subfigure{
    \begin{minipage}{5cm}
    \centering
    \includegraphics[width=5cm]{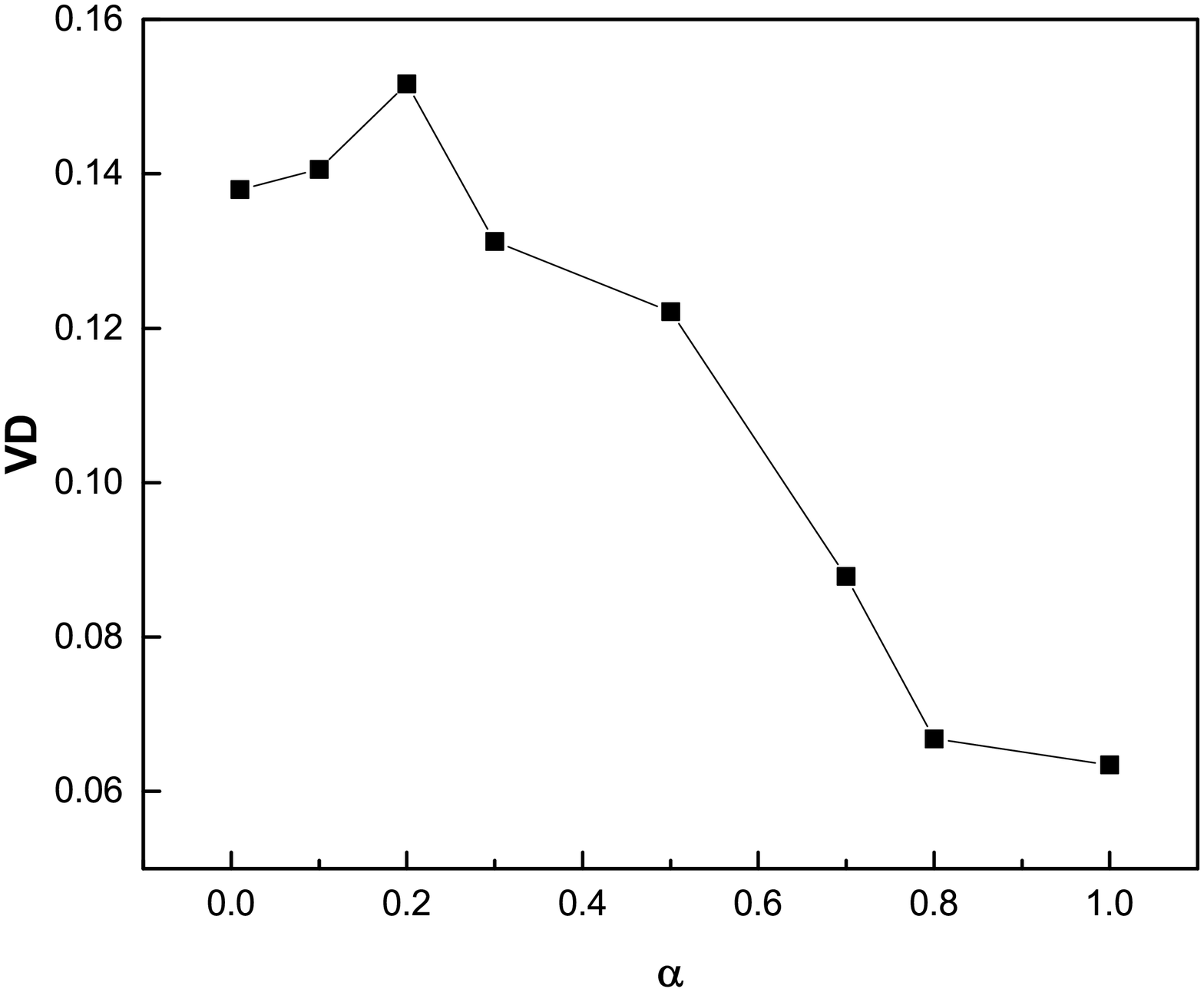}
    \end{minipage}}
     \caption{(Color online) The strength of interactions for l=0, 1, 2, which corresponds to s-, p-, and d-wave pairing, respectively.}\label{v1}
\end{figure*}
\begin{figure*}
\centering
\subfigure{
    \begin{minipage}{5cm}
    \centering
    \includegraphics[width=5cm]{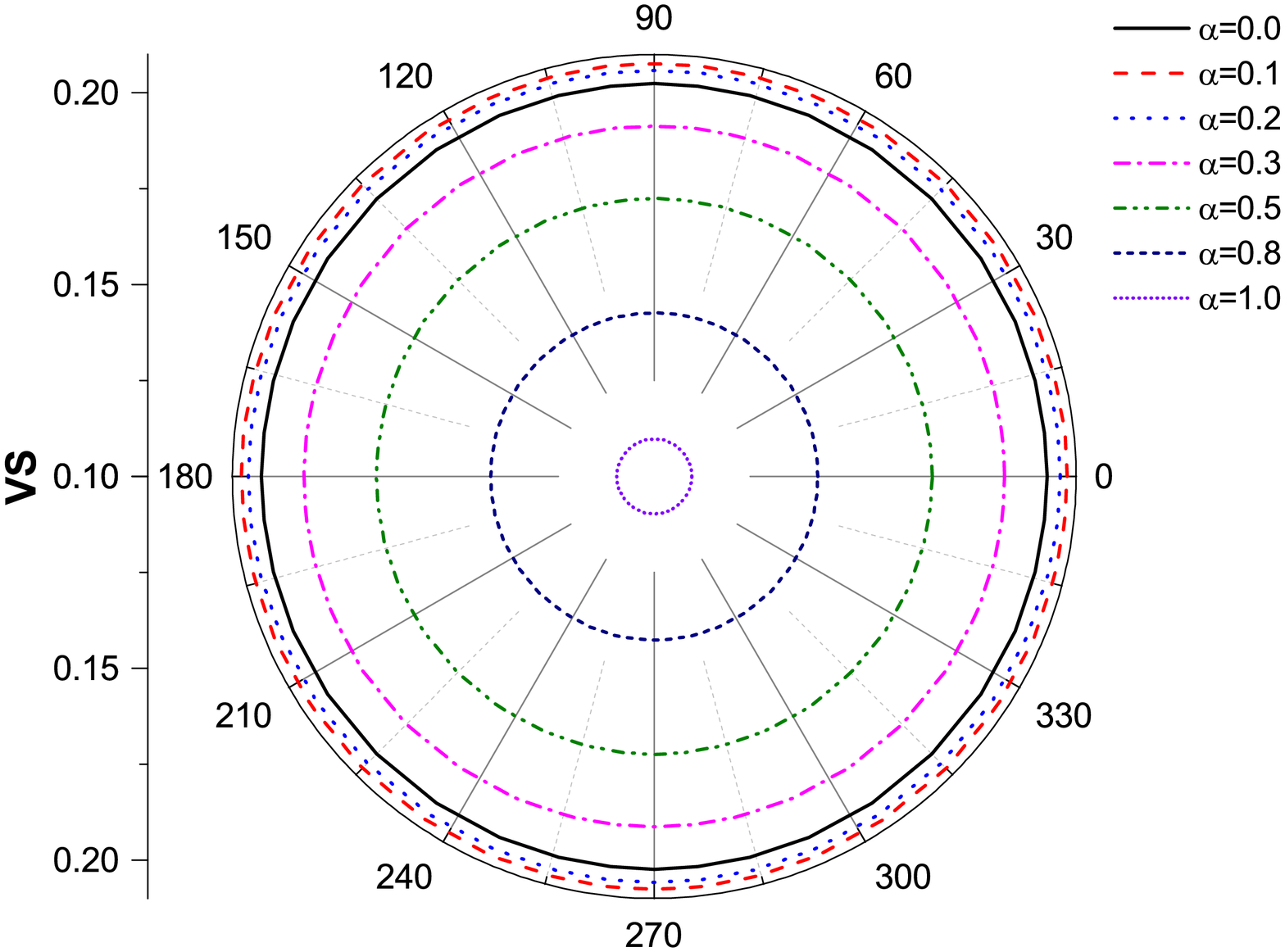}
    \end{minipage}}
\subfigure{
    \begin{minipage}{5cm}
    \centering
    \includegraphics[width=5cm]{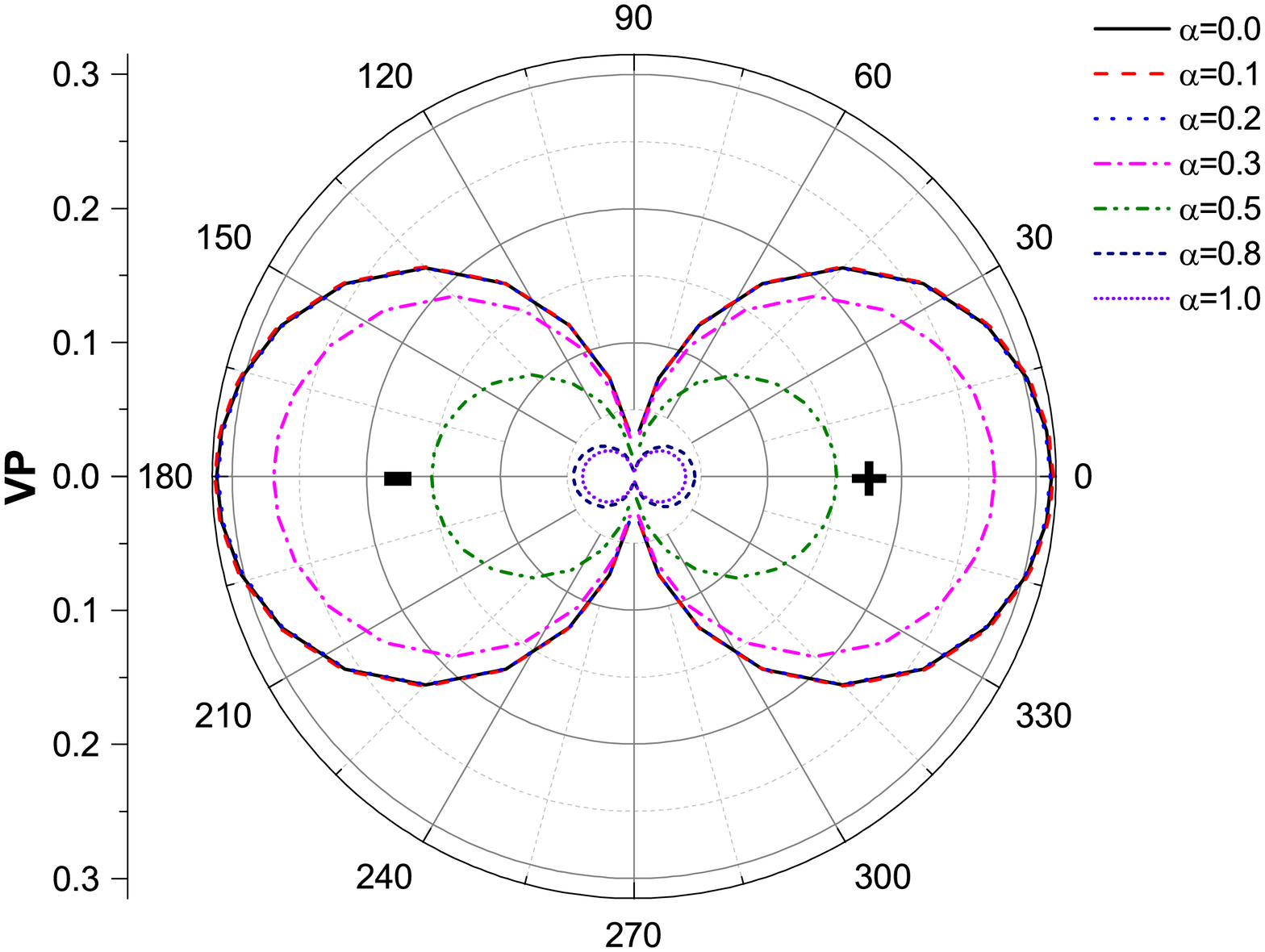}
    \end{minipage}}
\subfigure{
    \begin{minipage}{5cm}
    \centering
    \includegraphics[width=5cm]{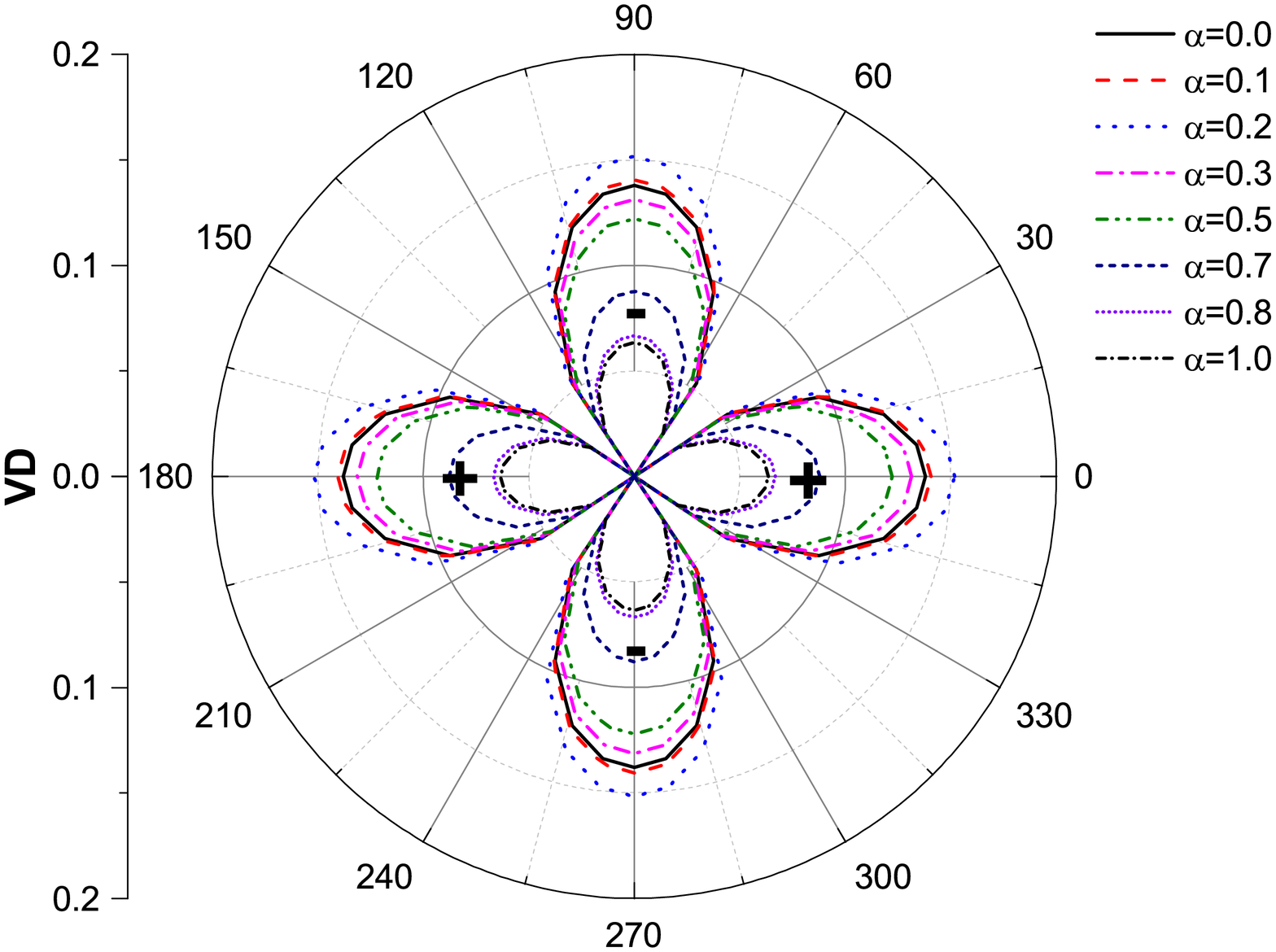}
    \end{minipage}}
     \caption{(Color online) The dependence of interactions on the angle of momentum for l=0, 1, 2, which corresponds to VS, VP, and VD, respectively.}\label{v2}
\end{figure*}
It is well known that the symmetry of pairs can be shown by the energy gap of superconductors, $\Delta_{SC}$, defined by BCS theory as,
\begin{align}
\Delta_{SC}(k)=\sum_{k^\prime}V(k^\prime,k)\frac{\Delta_{SC}(k^\prime)}{\sqrt{E_k+\Delta_{SC}(k^\prime)}}~.\label{gapeq}
\end{align}
Based on Eq.~(\ref{gapeq}), we can find that the symmetry of $\Delta_{SC}(k)$ is the same as the interaction on $V(k)$. In our model, the function $\Delta_{SC}(k)$ cannot be known directly, so the self-consistent equation is too difficult to solve. And thus we can discuss the symmetry of gap according to the symmetry of the interaction. To find out dominant channels of interaction, the $V_z$ is expanded in partial waves\cite{Fay1980,Monod1986}. For three dimensions, the interaction can be expressed in terms of spherical harmonics function $Y_l^m(x)$,
\begin{align}
&V(k,k^\prime)=\sum_{l,m}(2l+1)V_l^m Y_l^m(\theta_{k},\phi_{k})Y_l^m(\theta_{k^\prime},\phi_{k^\prime})~,\nonumber\\
&V_l^m=\int d\theta d\phi V(k,k^\prime)Y_l^m(\theta_{k},\phi_{k})Y_l^m(\theta_{k^\prime},\phi_{k^\prime})~.
\end{align}
where, $\theta$ and $\phi$ are solid angles of $\overrightarrow{k}$. Our model is two dimensional, so we have to use Legendre Polynomials $P_l(x)$ to express the interaction.
\begin{align}
&V(k,k^\prime)=\sum_{l}(2l+1)V_l P_l(\cos\theta_{k})P_l(\cos \theta_{k^\prime})~,\nonumber\\
&V_l=\int_{-1}^1 d\cos\theta_{k}d\cos\theta_{k^\prime} V(k,k^\prime)P_l(\cos\theta_{k})P_l(\cos \theta_{k^\prime})~.
\end{align}
where, $\cos \theta=k_x/\sqrt{k_x^2+k_y^2}$.
According to the BCS-type Hamiltonian, the interactions can be obtained as follow,
\begin{align}
H_{int}=&\sum_{k,k^\prime}V^S(k,k^\prime)c^{\dagger}_{k^\prime,\uparrow}c^{\dagger}_{-k^\prime,\downarrow}c_{-k,\downarrow}c_{k,\uparrow}\nonumber\\
&+\sum_{k,k^\prime,\sigma}V^T(k,k^\prime)c^{\dagger}_{k^\prime,\sigma}c^{\dagger}_{-k^\prime,\sigma}c_{-k,\sigma}c_{k,\sigma}~.
\end{align}
The spin-singlet interaction $V_l^S$ and triplet $V_l^T$, which consist of the interactions arising from the charge, longitudinal and transverse spin fluctuations, can be obtained by,
\begin{align}
V^S(k,k^\prime)&=V^c(k,k^\prime)+V^z(k,k^\prime)-2V^{+-}(k,k^\prime)~,\nonumber\\
V^T(k,k^\prime)&=V^c(k,k^\prime)-V^z(k,k^\prime)~.
\end{align}

To study the effect of SOC on $V_l$, we numerically calculated the strength of $l=0, 1, 2$, that is s-, p- and d-wave channel of interaction, with the strength of SOC $\alpha$=0.01, 0.1, 0.2, 0.3, 0.5, 0.8, 1.0. The relation of the strength of partial wave for interaction to SOC has been shown in Fig.~\ref{v1}. We can find that the values of $l=0$ and $l=1$ are negative and $l=2$ is positive. In order to facilitate comparison, the results without SOC are also calculated. The strength of s-wave potential, $l=0$, is about -0.17, and p-wave, $l=1$, is position and very small, 0.05. For $l=2$, d-wave, the strength is about -0.14. It means that SOC is in favor of s- and p-wave attractive interactions mediated by AFM fluctuations other than d-wave, which is different from the case without SOC\cite{Schrieffer1989,Anderson2016}, where d-wave pairing is dominant. Meanwhile, the strength of p-wave potential is as strong as s-wave. So the spin-orbit coupling could bring out the mixture of spin-singlet and spin-triplet Cooper pairs and the orbital degree of freedom is an admixture of s+p-wave. 
As mentioned in the introduction, SOC could lead to the mixture of spin-singlet and spin-triplet Cooper pairs\cite{Gorkov2001}. For spin-singlet Cooper pairs, the space wave function should be symmetric, for example, s-wave or d-wave, and for triplet, the space wave function should be antisymmetric, e.g. p-wave or f-wave. Our calculations suggest that s+p-wave Cooper pair be favorable with respect to the model considered in our works.It may be helpful in understanding the pairing symmetry of NCS superconductors. Many works on NCS superconductors have reported that the spin degree of freedom is the mixture of spin-singlet and triplet\cite{Yuan2006,Goryo2012,Shigeta2013}, however a consensus on the symmetry of orbital degree of freedom has not been reached. According to our calculations, SOC tends to form the s+p-wave Cooper pairs which is mediated by AFM fluctuations when the Hubbard model is adopted, which is agreed with Ref.~[\onlinecite{Tada2008}] and Ref.~[\onlinecite{Yanase2008}]. In contrast to the two papers, we calculate the partial waves of the effective interaction to study the pairing symmetry, which is more direct than them. Ref.~[\onlinecite{Yokoyama2007}] and [\onlinecite{Goryo2012}] suggest that SOC should induce d+f or s+f pairing states, but their models are different from ours. For the two-band models, some papers also indicate that SOC could play an important role in the p-wave pairing state. Sigrist {\it et al.}\cite{Sigrist2000} and Annet {\it et al.}\cite{Annett2006} have studied t-J model and an attractive Hubbard model, respectively. Both of them stated that the chiral p-wave state of Sr$_2$RuO$_4$ should be due to SOC. It seems that the effect of SOC on pairing symmetry might be dependent on the individual models.

For the s- and p-wave, the strength of interaction decreases with the increasing of SOC when the strength is strong. It implies that the large SOC might be bad for the superconductivity. However, the interaction increases initially with SOC and decreases afterwards. It seems that the SOC suppresses the magnetization, which possibly enhance the spin fluctuations. Furthermore, the interaction induced by spin fluctuations is promoted. When the strength of SOC is very large, all the pairing potentials are suppressed by SOC. It indicates that large SOC is bad for superconductivity which is agreed with the dependence of critical temperature on SOC\cite{Grzybowska2017}. Fig.~\ref{v2} illustrate the dependence of interactions on the momentum. According to Eq.~(\ref{gapeq}), it could describe the symmetry of the gap of superconductivity. Obviously, the s-wave is angular-isotropy, and it may be a conventional s-wave state. For p-wave, the state should be $l=1$ and $m=0$ in terms of the Legendre function of two dimension. And d-wave potential is also m=0, which is the same as p-wave.

\section*{V. Conclusion}
In summary, we have studied the ground state of the two-dimensional Hubbard model with Rashba SOC on a square lattice. Both the results obtained by MFA and RPA show that the sublattice magnetization decreases with the increasing of SOC for a fixed Hubbard interaction. Moreover, the magnetization for RPA is smaller than MFA with the same U and $\alpha$. The suppression of AFM order caused by SOC might be resulted from that SOC broadens the sub-Hubbard bands. Besides, a gapped energy spectrum of transverse spin fluctuations and an anisotropy of spin susceptibility, which are brought about by SOC, are present.

Furthermore, we have discussed the effective pairing interactions between electrons mediated by AFM spin fluctuations in the case of weak hole doping. The calculations about the partial waves of interactions indicate that p-wave potential can be induced by SOC. The d-wave potential which is dominant without SOC is suppressed by SOC. Moreover, the strength of s-wave always exists whether SOC is present or not. It seems that the SOC tends to form s+p paring rather than s+d pairing.

$\textit{Note added}.$ We are just aware of a numerical work studying the mechanism of p-wave Cooper pairs\cite{Kim2017}. They state that the degeneracy of various p-wave states is split by the magnetic anisotropy. The anisotropy might result from SOC. This paper indicates that the symmetry of Cooper pairs mediated by spin-fluctuations is still a hot topic.
\begin{acknowledgments}
This work is supported by the National Natural Science Foundation of China (Grant No. 11274039) and the Fundamental Research Funds for the Central Universities of China.
\end{acknowledgments}

\end{document}